\documentclass[a4paper,fleqn]{cas-sc}

\usepackage[numbers]{natbib}
\usepackage{lineno}
\usepackage{subfig}
\usepackage{booktabs}
\usepackage{amsmath,amssymb}
\usepackage[export]{adjustbox}

\def\tsc#1{\csdef{#1}{\textsc{\lowercase{#1}}\xspace}}
\tsc{WGM}
\tsc{QE}

\begin{document}
\let\WriteBookmarks\relax
\def\floatpagepagefraction{1}
\def\textpagefraction{.001}

\shorttitle{A MT-BCA-CNN Model for Few-shot UATR}   

\shortauthors{Huang et al.}

\title[mode = title]{A Multi-task Learning Balanced Attention Convolutional Neural Network Model for Few-shot Underwater Acoustic Target Recognition}  

\tnotemark[1]

\tnotetext[1]{This work was financially supported in part by the National Natural Science Foundation of China (42404001), in part by the National Key Research and Development Program of China (2024YFB3909701), in part by the National Natural Science Foundation of China (62271459), in part by Natural Science Foundation of Shandong Province (ZR2023QF128).}

%

















\author[1]{Wei Huang}[type=author,
    auid=000,bioid=1,
    orcid=0000-0002-8284-2310]
\fnmark[1] 
\ead{hw@ouc.edu.cn} 
\credit{Conceptualization of this study, Methodology, Original draft preparation}

\author[1]{Shumeng Sun}
\fnmark[1] 
\credit{Data curation, Software, Original draft preparation}
\author[1]{Junpeng Lu}
\fnmark[1] 
\credit{Data curation, Software}
\author[1]{Zhenpeng Xu}
\fnmark[1] 
\credit{Data curation, Software}
\author[1]{Zhengyang Xiu}
\fnmark[1] 
\credit{Data curation, Software}
\author[1]{Hao Zhang}
\fnmark[1] 
\credit{Conceptualization of this study, Methodology, Original draft preparation}
\cormark[1] 
\ead{zhanghao@ouc.edu.cn}
\cortext[1]{Corresponding author, Wei Huang and Shumeng Sun contributed equally to this work.} 
\address[1]{Faculty of Information Science and Engineering, Ocean University of China, Qingdao 266100, China}

\begin{abstract}
Underwater acoustic target recognition (UATR) is of great significance for the protection of marine diversity and national defense security. The development of deep learning provides new opportunities for UATR, but faces challenges brought by the scarcity of reference samples and complex environmental interference. To address these issues, we proposes a multi-task balanced channel attention convolutional neural network (MT-BCA-CNN). The method integrates a channel attention mechanism with a multi-task learning strategy, constructing a shared feature extractor and multi-task classifiers to jointly optimize target classification and feature reconstruction tasks. The channel attention mechanism dynamically enhances discriminative acoustic features such as harmonic structures while suppressing noise. Experiments on the Watkins Marine Life Dataset demonstrate that MT-BCA-CNN achieves 97\% classification accuracy and 95\% $F1$-score in 27-class few-shot scenarios, significantly outperforming traditional CNN and ACNN models, as well as popular state-of-the-art UATR methods. Ablation studies confirm the synergistic benefits of multi-task learning and attention mechanisms, while a dynamic weighting adjustment strategy effectively balances task contributions. This work provides an efficient solution for few-shot underwater acoustic recognition, advancing research in marine bioacoustics and sonar signal processing.
\end{abstract}


\begin{highlights}
\item A multi-task few-shot learning framework for underwater acoustic target recognition (UATR).
\item A channel-attention-assisted convolutional neural network model for UATR.
\item A combination of attention layer with Gaussian kernel to prevent over-concentration.
\end{highlights}


\maketitle

\section{Introduction}

Acoustic waves are widely recognized as the optimal information carrier for sensing and identifying underwater targets, being the only known form of energy that can propagate over long distances under water \cite{1}. Underwater acoustic target recognition (UATR) is of great significance in fields such as marine resource exploration, disaster warning, and marine biodiversity conservation. For example, in marine biological research, UATR technology can be employed to track individual or groups of marine organisms, which assists scientists in studying behavioral patterns, distribution, and migration of marine life, thereby providing a foundation for marine biodiversity conservation. In ocean environmental monitoring, the technology proves valuable for deploying and tracking the positioning of detection equipment during the observation of submarine volcanic activity, seabed landslides, and other geological hazards. To better advance development in related fields, exploring more precise UATR methods has emerged as a focal point for researchers. 

\indent In recent years, with the development of theoretical research in deep learning, its applications in computer vision, audio processing, data augmentation, and noise reduction have achieved remarkable success. This has provided novel solutions for research directions in hydroacoustic engineering, such as UATR, audio classification, and object detection \cite{2}. However, the unique underwater environment still poses severe challenges for UATR. The velocity of sound in the ocean usually exhibits a layered structure in the vertical direction, leading to acoustic refraction, which may cause distortion of sonar signals and increase the difficulty of target recognition. Meanwhile, the marine environment contains noise from various sources, such as biological activity, shipping, and natural turbulence. These noises can mask the weak signals reflected by the target, interfere with the detection of acoustic characteristics, reduce the accuracy of feature extraction, and ultimately affect the reliability of target recognition. Moreover, during underwater propagation, acoustic waves generate multiple propagation paths due to reflections at boundaries such as the sea surface, seabed, and heterogeneous media within the water column. The multipath effects cause temporal and spatial spreading of received target signals, degrading the performance of beamforming-based target recognition methods. At last, the expensive economic costs of underwater equipment, and difficult maintenance conditions lead to a scarcity of data samples, which poses difficulties for studying the characteristics of underwater targets. In a word, the combination of high ambient noise, strong interference, complex underwater acoustic channels, and limited data availability poses significant challenges to UATR \cite{3}.

\indent Traditional audio processing techniques are grounded in foundational methods such as time-domain analysis\cite{4}, frequency-domain analysis\cite{5}, and cepstral analysis\cite{6}, where time-domain analysis is one of the most fundamental approaches in audio processing. By examining variations in audio signals over the time dimension, key features such as  signal duration and zero-crossing rate can be extracted. For instance, sounds emitted by certain fish species exhibit  specific rhythmic patterns and temporal duration characteristics in the time domain. Such features can serve as distinguishing criteria for identifying different fish species. The zero-crossing rate, which quantifies how frequently a signal alternates between positive and negative values, may reveal unique patterns in species with distinct vocalization styles. These characteristics could be leveraged in audio classification tasks for underwater biological studies. Frequency-domain analysis techniques employ methods such as Fourier transform to convert audio signals from the time domain to the frequency domain, enabling the analysis of signal frequency components. Different underwater organisms produce sounds within distinct frequency ranges. By examining spectral features, critical information for distinguishing marine species can be extracted. Commonly used spectral characteristics include spectral bandwidth, energy distribution across frequencies, and spectral peak locations. Cepstral analysis is obtained by applying an inverse Fourier transform to the logarithmic power spectrum of an audio signal. The Mel-Frequency Cepstral Coefficients (MFCCs), a widely used form of cepstral features, are notable for simulating the human auditory system's perception of sound frequencies. This characteristic makes MFCCs particularly valuable in feature extraction processes for underwater biological audio analysis. In the context of underwater acoustics, MFCCs are similarly leveraged to capture perceptual frequency characteristics, though adaptations may be required to account for unique environmental factors such as low-frequency dominance in marine sounds or ambient noise interference. In 2022, Abdul et al. reviewed the application of MFCC combined with deep learning models in automatic speech recognition (ASR) to enhance recognition performance \cite{6}. Later, Tang et al. proposed a novel method to convert Mel-spectrograms into three-dimensional data and introduced an efficient 3D Spectrogram Network (3DSNet) that independently processes temporal and frequency dimensions, significantly improving feature representation capabilities \cite{8}. Other feature extraction methods in UATR also include spectral peak analysis, wavelet transforms, and Linear Predictive Cepstral Coefficients (LPCCs), and so on. 

\indent With advancements in underwater acoustics research, many audio classification algorithms driven by machine learning have emerged. Early-stage methods included the Hidden Markov Model (HMM) for temporal sequence modeling and the Gaussian Mixture Model (GMM) for probabilistic distribution modeling of acoustic features. The Hidden Markov Model (HMM) is grounded in probabilistic statistical theory and is used to model temporally sequential underwater biological sounds, effectively capturing the temporal evolution of acoustic signals. In 2018, Mohammed et al. proposed an underwater target classifier based on Gamma-Tone Cepstral Coefficients (GTCCs) and a Hidden Markov Model (HMM) with 20 states, which was evaluated using real-world field data \cite{9}. However, the parameter estimation process of Hidden Markov Models (HMMs) is computationally complex and requires a large number of data samples, making it challenging to meet practical requirements in real-world audio classification tasks. In the same year, Xue and Jiang proposed an adaptive algorithm based on the Gaussian Mixture Model (GMM) to investigate target detection and classification using Ultra-Wideband (UWB) signals under varying weather conditions \cite{10}. Compared with HMM, GMM offers flexibility in modeling complex data distributions by assuming that the data is generated from a mixture of multiple Gaussian distributions. This allows GMM to effectively capture the intricate frequency and amplitude characteristics of diverse underwater biological sounds. Nevertheless, for large-scale underwater acoustic datasets, the training of GMM can be time-consuming, and its computational overhead may hinder real-time classification performance. 

\indent Bist and Singh proposed a comprehensive analysis of Support Vector Machine (SVM) variants and their optimization techniques to investigate key challenges in model selection, computational cost-effectiveness, and algorithm optimization for SVMs \cite{11}. The Support Vector Machine (SVM) is a supervised learning algorithm whose core idea is to identify an optimal hyperplane that separates data of different classes while maximizing the margin. However, SVM was originally designed for binary classification problems and is difficult to handle multi classification problems. The structure of a decision tree resembles a flowchart, naturally handling audio data with multiple features by sequentially evaluating different attributes for classification. However, decision trees are prone to overfitting, and the local optimality of their splitting criteria can compromise classification performance. In contrast, the Random Forest algorithm addresses these limitations by constructing multiple decision trees through bootstrap aggregating (bagging), where each tree is trained on a randomly sampled subset of the data with replacement. This ensemble approach reduces misclassification errors caused by data noise or overfitting in individual trees, making it more robust for classifying complex and variable underwater bioacoustic data. Nevertheless, when processing high-dimensional features, Random Forests may demand substantial computational resources, resulting in prolonged training times. With the gradual evolution of statistical models in underwater audio classification, researchers have continuously explored new methods to enhance classification performance. However, traditional statistical approaches increasingly reveal limitations when handling complex underwater acoustic data, such as struggles with high-dimensional feature spaces, non-stationary noise interference, and nonlinear signal patterns.

\begin{figure}[!htbp]
	\centering
	\includegraphics[width=1\linewidth]{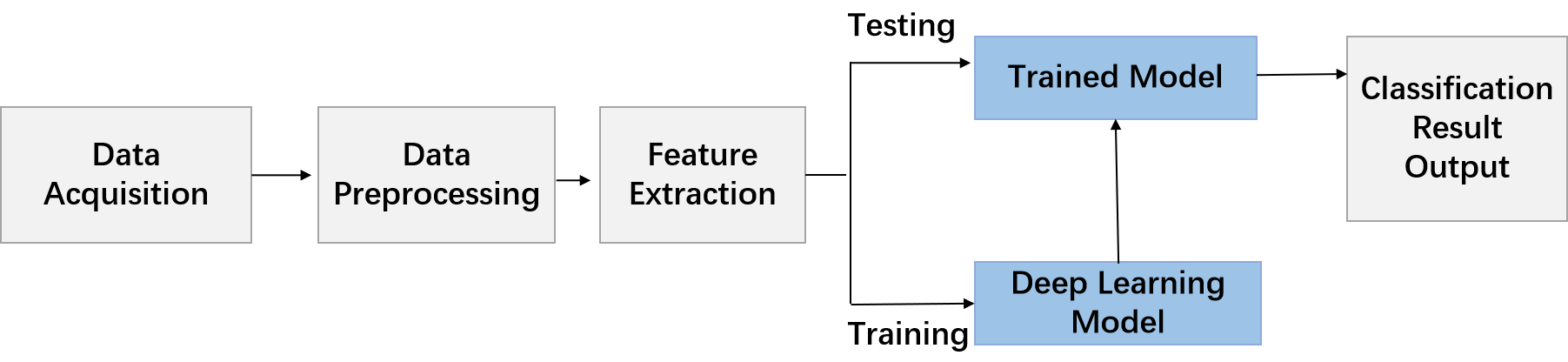}
	\caption{Target Recognition Task Process}
	\label{fig:1}
\end{figure}
\indent Against this backdrop, the emergence of deep learning technologies has brought new possibilities to underwater audio classification, marking a transformative shift in the field. The general deep-learning-based audio recognition process is illustrated in Figure\ref{fig:1}, which encompasses the following workflow: target signal acquisition, database creation, data preprocessing, feature extraction, and audio classification. In the field of deep learning, convolutional neural networks (CNNs) and recurrent neural networks (RNNs) are two dominant architectural paradigms. CNN, renowned for their spatial feature extraction capabilities in image processing, have also proven highly effective for audio signal analysis. They excel at processing two-dimensional spectral representations such as Lofar grams (low-frequency analysis and ranging spectrograms) and Mel-spectrograms, which encode audio signals into time-frequency grids. RNNs, on the other hand, specialize in capturing temporal dynamics within sequential data. This makes them particularly suited for analyzing one-dimensional feature sequences like MFCC trajectories, demodulation of envelope modulation on noise (DEMON) spectra or wavelet coefficients, which reflect time-varying patterns in acoustic signals.

\indent In 2018, Aodha et al. developed an open-source system that employed machine learning, specifically a supervised deep convolutional neural network (BatDetect CNN), to learn features directly from raw audio data for automated detection of bat echolocation calls during the search phase in noisy real-world recordings \cite{12}. In 2019, Yang et al. proposed a novel end-to-end neural network model—the Auditory Perception-Inspired Deep Convolutional Neural Network (ADCNN)—for underwater acoustic target recognition (UATR). This model simultaneously addressed the decomposition, feature extraction, and classification of ship-radiated noise, achieving integrated processing for marine target recognition tasks \cite{13}. In 2020, Miao et al. designed a deep convolutional neural network architecture tailored for underwater acoustic (UWA) signal classification, termed the TF-Feature Network (TFFNet). To reduce network complexity while integrating contextual information across multiple scales, they proposed an Effective Feature Pyramid (EFP) forward feature fusion method. The performance of TFFNet with EFP and sparse activation (ACT) was evaluated on two UWA datasets, demonstrating superior efficiency and accuracy in noisy environments \cite{14}. In 2021, Panetta et al. introduced a Cascaded Residual Network for Underwater Image Enhancement (CRN-UIE), leveraging generative adversarial networks (GANs) to transform degraded underwater visuals into enhanced/clearer representations \cite{15}. This method aimed to improve downstream tasks like object tracking by mitigating underwater-specific distortions (e.g., light scattering, color casts). In 2021, Liu et al. proposed a CRNN model for underwater target recognition, involving three steps: extracting three channels from the log-Mel spectrogram as input, applying data augmentation to expand the dataset time and time-frequency domains, and leveraging the CRNN model for automatic feature learning and target classification \cite{16}. In 2021, Xu et al. proposed an Attention-Based Neural Network (ABNN) for target recognition in multi-source interference pressure spectrograms, utilizing attention modules to explore the internal mechanisms of neural networks \cite{17}. In 2023, Christopher et al. introduced the ORCA-SPY framework for fully automated sound source simulation, classification, and localization in passive killer whale acoustic monitoring. The framework integrates a ResNet-based convolutional neural network (CNN) called ANIMAL-SPOT for audio segmentation and PAMGuard’s TDOA-based localization plugin. Validated experiments demonstrated ORCA-SPY’s capability to detect and track target signals even in noisy environments \cite{18}. In 2023, Xue et al. proposed the CamResNet model by enhancing the ResNet architecture with a channel attention mechanism, which strengthens stable spectral features and mitigates unstable signals caused by Doppler shifts \cite{19}. In the same year, Feng et al. introduced the UATR-Transformer, a Transformer-based model for underwater acoustic target recognition, employing Mel-filter bank (Mel-fbank) features as input representations. Experimental validation confirmed the model’s strong performance in complex marine environments \cite{20}. 

\indent Above model methodologies predominantly rely on extensive labeled datasets for implementation, overlooking the practical challenges of acquiring large-scale data samples in real-world marine environments. Consequently, while these methods demonstrate exceptional performance under ideal conditions, they exhibit notable shortcomings in addressing few-shot learning scenarios. To tackle these issues, Abhinav et al. addressed the degradation of speech recognition performance in noisy environments by proposing a multi-task learning-based audio-visual speech recognition (MTL-DNN) method. By simultaneously learning audio-visual fusion features and visual feature mapping, the model demonstrated significant performance improvements under high-noise conditions \cite{21}. In 2019, Kong et al. introduced a cross-task learning framework for audio tagging, sound event detection, and spatial localization, aiming to provide a unified baseline system for multiple audio recognition tasks in the DCASE 2019 Challenge \cite{22}. Apoorv proposed a cross-domain MTL framework for joint optimization of critical downstream computer vision tasks—object detection and saliency estimation, in scenarios where no co-annotated datasets exist \cite{23}. In 2023, Yan et al. developed an MTL-driven Acoustic Scene Classification (ASC) method, leveraging topic-based soft labels and a mutual attention mechanism to refine feature representations and boost classification accuracy in diverse acoustic scenes \cite{24}. However, few-shot learning is still receiving less attention in the field of UATR.

To address the challenges of few-shot learning issue, multipath effect and noise interference, we explore an innovative methodology for UATR by establishing a convolutional neural network model augmented with channel attention mechanisms and establish a multi-task learning training framework for UATR, which is termed as the multi-task balanced attention convolutional neural network (MT-BCA-CNN). This approach not only maximizes the utilization of limited labeled data but also enhances the model’s capability to recognize complex acoustic signals and improve generalization performance through the synergistic integration of multi-task learning and channel attention mechanisms. The primary contributions of this work are summarized as follows: 
\begin{enumerate}
	\itemsep=0pt
	\item To enable the model to filter out interference from complex noise, focus on critical regions, and visualize the most influential areas in decision-making—thereby enhancing interpretability—we propose an attention mechanism-based underwater bioacoustic classification model. By integrating an attention mechanism, the model rapidly localizes key regional features in noisy samples, significantly improving classification accuracy.
	\item Given the practical challenges of acquiring sufficient labeled samples and the performance degradation of existing models in few-shot scenarios, we establish a multi-task learning training framework with an attention mechanism to enable rapid adaptation to new classification tasks. By employing shared feature extractors and task-specific classifiers, the model simultaneously learns classification and feature reconstruction tasks, thereby enhancing generalization capability and robustness. This approach is particularly critical in few-shot learning contexts, as it maximizes the utility of limited labeled data through joint optimization of multiple tasks, significantly improving recognition accuracy for diverse underwater acoustic targets. 
	\item To prevent over-concentration, we propose a combination architecture of an attention layer and a Gaussian kernel, which smooths the attention weights and mitigates abrupt feature transitions.
\end{enumerate}

The rest of this paper is organized as follows. The proposed method is proposed in Section 2. Results and discussions are given in Section 3, and the conclusions are drawn in Section 4.

\section{Methodology}
In this paper, we propose a few-shot UATR method, centered on a convolutional neural network model that integrates multi-task learning with a channel attention mechanism. Below, we elaborate on the architecture and implementation details of the proposed MT-BCA-CNN, including the design of the attention module and the multi-task learning training strategy.
\subsection{Overview of Model Architecture}
\begin{figure}
	\centering
	\includegraphics[width=1\linewidth]{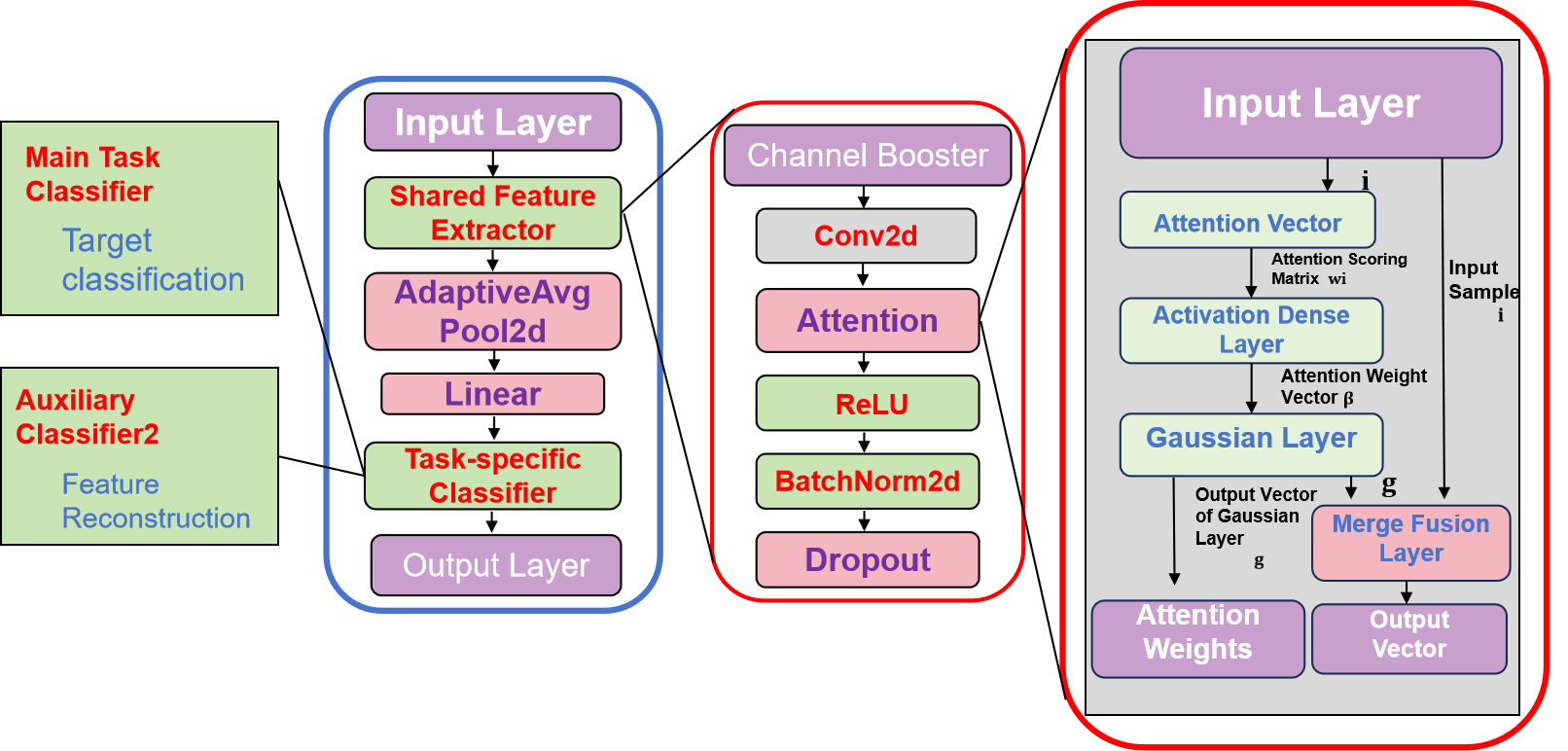}
	\caption{MT-BCA-CNN Model Architecture}
	\label{fig:2}
\end{figure}
The MT-BCA-CNN network model proposed in this study is a convolutional neural network that integrates a channel attention mechanism under multi-task learning\cite{25} framework, comprising an input layer, shared feature extractor, adaptive average pooling layer, fully connected layers, task-specific classifiers, and an output layer. As illustrated in Figure \ref{fig:2}, the multi-task learning module consists of a shared feature extractor and two task-specific classifiers. The shared feature extractor is a channel attention mechanism-based CNN that hierarchically extracts features through four convolutional blocks. Each block contains five sequential layers: a channel enhancement module, a convolutional layer, an attention module, a ReLU activation layer, and a Batch Normalization layer (BatchNorm2D).

During processing, input data first passes through the convolutional layer to extract audio features. These features are then refined by the attention module, where channel enhancement generates attention weights to dynamically amplify critical channel-wise features in the audio feature maps while suppressing less relevant ones. After the attention module, the data passes through a ReLU activation layer to introduce non-linearity, followed by a batch normalization layer for standardized output. In the first layer of the convolutional block, there are 2 input channels, 8 output channels, 3×3 kernel. The output spatial dimensions (height and width) correspond to the time steps (t) and frequency bins (f) of the audio feature map. Subsequent layers follow the same processing sequence (convolution → attention → ReLU → BatchNorm):
\begin{itemize}
	\item Conv2 (Layer 2): 8 input channels, 16 output channels.
	\item Conv3 (Layer 3): 16 input channels, 32 output channels.
	\item Conv4 (Layer 4): 32 input channels, 64 output channels.
\end{itemize}

After traversing the convolutional block, the data enters an adaptive average pooling layer to reduce the spatial dimensions of the feature maps. Then, the feature maps from the pooling layer are flattened, and the data are mapped to task-specific classifiers via a fully connected (FC) layer. The model includes a primary task classifier for target classification and an auxiliary classifier for feature reconstruction. The primary classifier uses the FC layer to map the feature vector to the class probability space, outputting a probability distribution over target categories. The auxiliary classifier employs transposed convolutional layers to upsample the shared feature vector into a time-frequency representation matching the original input, minimizing reconstruction error to enhance feature representation and improve robustness to noise. This multi-task design enables joint optimization, leveraging both discriminative classification and noise-invariant feature learning for superior performance in data-scarce underwater acoustic scenarios.

Our model hierarchically extracts audio features through multiple convolutional layers and enhances focus on critical patterns via an attention module. Finally, classification is performed through fully connected layers and multi-task learning, with Dropout applied to improve generalization and prevent overfitting.

\subsection{Attention Module}
In the MT-BCA-CNN architecture shown in Figure~\ref{fig:2}, the attention module is positioned after the convolutional layers. Within this module, input features are weighted by an activation-based dense layer to modulate the sensitivity of frequency components. To prevent over-concentration, the attention layer is followed by multiplication with a Gaussian kernel, which smooths the attention weights and mitigates abrupt feature transitions. For underwater biological signals, beyond continuous broadband line spectra, this design reduces the impact of Doppler shifts, measurement errors, and other factors causing random variations in features. The attention module then combines the Gaussian-smoothed output sequence with the input sequence through a merge layer, effectively acting as a mask that retains only target-relevant features while suppressing noise and interference. 

To construct an attention module using a soft attention masking mechanism, the most fundamental soft masking layer can be implemented through a dense layer activated by softmax. Each input data sample $i$ is a vector of length $f$, and the attention weight vector $\beta$ of size $f$ is expressed as:
\begin{equation}\beta = \text{softmax}(W_i),\end{equation}
where, $W_i$ is an attention scoring matrix of size $f$, learned by the CNN and softmax function, which quantifies the relevance between each input feature in $i$ and the target event. Higher relevance corresponds higher score. Subsequently, the Gaussian layer is defined by multiplying the above result with a Gaussian kernel function, expressed as a matrix operation:
\begin{equation}
	K = \left( -\frac{1}{2\beta^2} 
	\begin{bmatrix}
		\frac{k_{11}^2}{2} & \frac{k_{12}^2}{2} & \cdots & \frac{k_{1S}^2}{2} \\
		\frac{k_{21}^2}{2} & \frac{k_{22}^2}{2} & \cdots & \cdots \\
		\vdots & \vdots & \ddots & \vdots \\
		\frac{k_{S1}^2}{2} & \cdots & \cdots & \frac{k_{SS}^2}{2}
	\end{bmatrix}
	\right), \quad k_{mj} = j - m, \quad m, j \in [0, S], 
\end{equation}
\begin{equation}
	g = \frac{1}{\omega\sqrt{2\pi}} \exp\left( -\frac{1}{2\omega^2} K \right) \beta.
\end{equation}
The Gaussian layer, functioning as the attention weight layer, produces an output vector $g$ of size $f$, where $\omega$ represents the Gaussian function's standard deviation. A higher $\omega$ value broadens the attention distribution, reducing overemphasis on isolated frequency components and improving resilience to spectral variations (e.g., Doppler effects or measurement noise). The attention module’s final output $o$, an $f$-size vector, is derived via the Hadamard product ($\odot$):
\begin{equation}
	o = g \odot i.
\end{equation}

The channel attention module in this work consists of two parallel branches: an average pooling branch and a max pooling branch \cite{26,27}. The average pooling branch extracts global average features from the input feature map via a global average pooling operation, while the max pooling branch captures global maximum features through a global max pooling operation. These two branches collectively capture distinct statistical characteristics of the feature map, enabling complementary feature enhancement for robust acoustic target recognition. The pooled features are fed into a fully connected layer \cite{28} to generate channel-wise attention weights. The final attention weights are obtained by applying a sigmoid function to the sum of the outputs from both branches. These weights recalibrate the importance of each channel in the feature map, enhancing the model's focus on critical channels (e.g., harmonic-rich frequency bands in cetacean vocalizations). The data undergoes multi-layer convolutional operations to progressively extract low-level to high-level features. Each convolutional layer is followed by a ReLU activation function and BatchNorm2d normalization, introducing non-linearity and stabilizing the training process.

\subsection{Multi-Task Learning Structure}
MTL is a machine learning framework that improves data efficiency and model generalization by jointly optimizing multiple related tasks, exploiting shared representations while preserving task-specific distinctions. This method is especially vital in few-shot  underwater acoustic recognition, where the scarcity of labeled data necessitates cross-task knowledge sharing to reduce overfitting and enhance resilience to noise (e.g., ship noise or wave activity).
\begin{figure}
	\centering
	\includegraphics[width=1\linewidth]{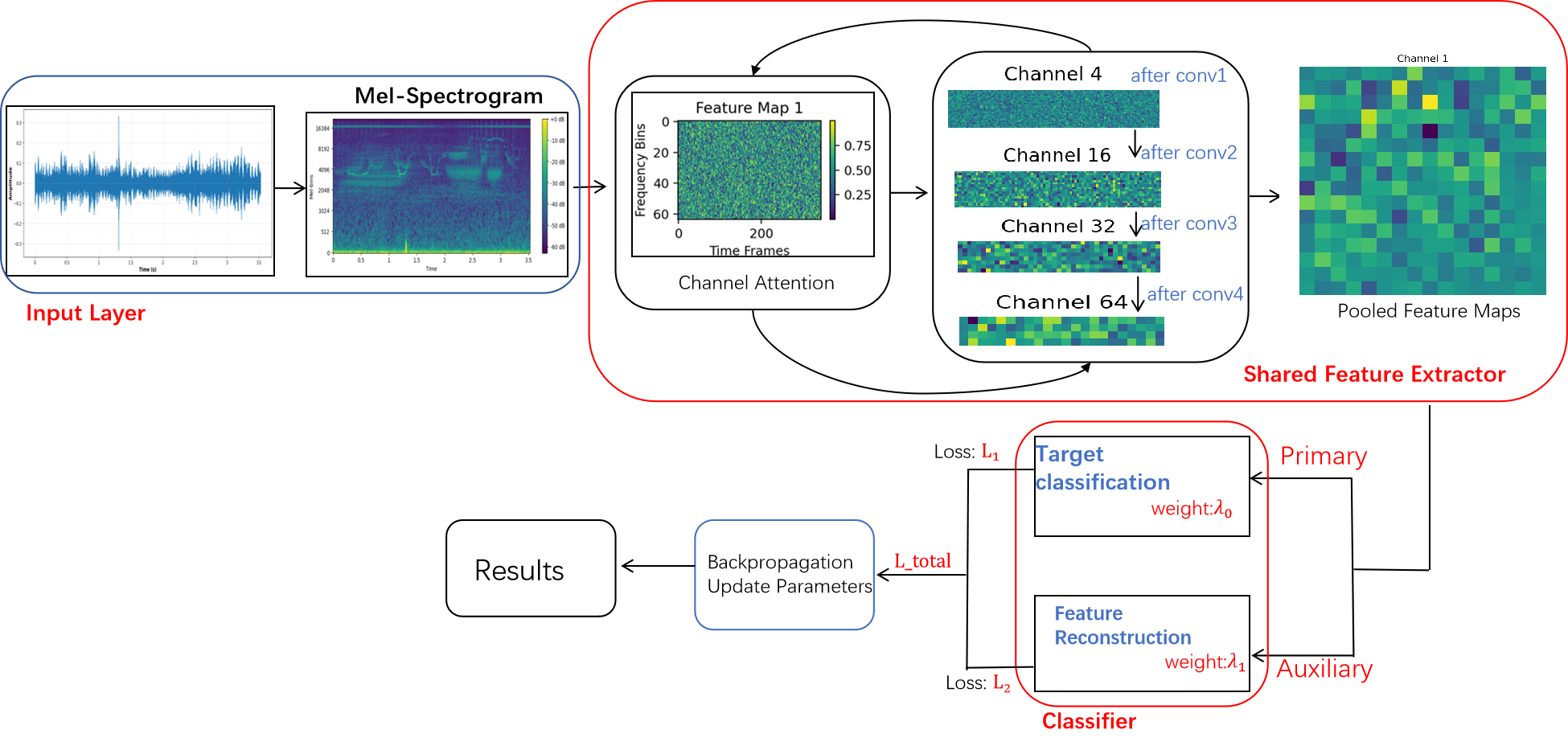}
	\caption{Flowchart of multi-task learning implementation, where $\lambda_0$ and $\lambda_1$ denote the weights of the task-specific classifiers, $\ L_1$ and $\ L_2$ represent the task losses, and $L_{total}$ is the joint loss function}
	\label{fig:3}
\end{figure}
\subsubsection{Shared Feature Extractor}
The multi-task learning module in the proposed MT-BCA-CNN model comprises a shared feature extractor and task-specific classifiers. As shown in Figure \ref{fig:3}, the shared feature extractor serves as the module’s core component, implemented via a convolutional neural network (CNN) integrated with a channel attention mechanism \cite{29}. The input layer processes raw audio data through Mel-spectrogram feature extraction, converting it into a 2D numpy array. The extracted spectrogram is first refined by the channel attention mechanism using:
\begin{equation}
	aw = \sigma\Big( \mathrm{fc2}\big( \text{ReLU}( \mathrm{fc1}(\text{avg\_pool}(x)) + \mathrm{fc1}(\text{max\_pool}(x)) ) \big) \Big),
\end{equation}
\begin{equation}
	out = aw \odot x,
\end{equation}
where the input feature map $x$ has dimensions M$\times$N, and M and N denote height and width. The sigmoid function ($\sigma$) maps the fully connected layer’s output to the [0,1] range, generating channel attention weights. The input feature map $x$, with dimensions  M$\times$N (height and width), undergoes global average pooling $ avg\_pool(x)$ and global max pooling $max\_pool(x)$ to generate channel-wise statistics, producing an 1D vector of shape C$\times$1 (where C is the number of channels). These pooled features are then compressed into a low-dimensional space via the first fully connected layer $fc1$, and the sigmoid function $\sigma$ maps the output to the [0,1] range to compute channel attention weights, which highlight critical frequency components (e.g., harmonic signatures) while suppressing noise interference.The ReLU\cite{30} activation function introduces nonlinearity to the model. The second fully connected layer ($fc2$) subsequently maps the low-dimensional features back to the original channel dimension. Through these operations, attention weights ($aw$) are generated, which are then applied to the input feature map $x$ via element-wise multiplication, resulting in the refined output feature map that emphasizes task-relevant spectral patterns (e.g., transient pulses or tonal signals) while suppressing irrelevant noise.

The attention-processed feature map is fed into a convolutional layer for processing:
\begin{equation}
	x = \mathrm{BN}\big( \mathrm{ReLU}( \mathrm{Conv2d}(x) ) \big).
\end{equation}
The feature map from the convolutional layer is first processed using Formula (7),
it is then iteratively refined through channel attention mechanism 2 for 4 cycles. After the final convolutional layer, the feature map is fed into an adaptive average pooling layer:
\begin{equation}
	\boldsymbol{v} = \mathrm{AdaptAP}\big( \mathrm{ConvBlock}^4(x) \big) ,
\end{equation}
where $\boldsymbol{v}$ denotes the resulting pooled feature vector with a dimension of 64. This process extracts global features through average pooling and max pooling, followed by fully connected layers and a sigmoid function to generate attention weights, which adaptively recalibrate the importance of each channel in the input feature map. The data then undergoes multi-layer convolutional operations to progressively extract hierarchical features from low-level (e.g., edges, textures) to high-level (e.g., structural patterns). Each convolutional layer is coupled with a \cite{31}, enhancing nonlinear representational capacity and stabilizing the training dynamics by mitigating gradient-related issues (e.g., vanishing gradients, internal covariate shift).

Multi-task learning (MTL) leverages a shared feature extractor to enable more efficient learning of generalized feature representations from limited labeled data. By jointly optimizing across tasks, the model extracts high-level features from input acoustic signals that not only encapsulate task-critical information (e.g., target-specific spectral signatures) but also capture shared patterns across tasks (e.g., noise suppression mechanisms or harmonic structures common to multiple targets). This synergy enhances robustness and adaptability in data-scarce scenarios like underwater acoustic recognition.
\subsubsection{Task-Specific Classifiers}
The proposed MT-BCA-CNN model employs two task-specific classifiers. The feature vectors output by the shared feature extractor are fed into these two classifiers:
\begin{enumerate}[\textbullet]
	\item  The primary task classifier maps feature vectors to the target category space, generating probability distributions for each class.
	\item  The auxiliary task classifier  upsamples the shared feature vectors into time-frequency spectrograms of raw input signals via transposed convolutional layers, minimizing reconstruction errors to enhance feature representation.
\end{enumerate}
The joint loss function of the model is a weighted sum of the loss functions for each task. The primary task employs a cross-entropy loss function, defined as:
\begin{equation}
	L_1 = - \sum_{c=1}^C x_c \log(p_c), \label{eq1}
\end{equation}
where C is the total number of classes, $x_c$ represents the one-hot encoded ground-truth label, and $p_c$ is the predicted probability of the model assigning the input to class c. The reconstruction task utilizes the mean squared error (MSE) loss function, formulated as:
\begin{equation}
	L_2 = \frac{1}{N} \sum_{i=1}^N (y_{recon,i} - \hat{y}_{recon,i})^2,
\end{equation}
where N is the number of samples, $\hat{y}_{recon,i}$and $y_{recon,i}$ are the predicted and true values of signal features for the $i$th sample, respectively. The joint loss function is:
\begin{equation}
	L_{total} = \lambda_0 L_1 + \lambda_1 L_2 , \label{eq4}
\end{equation}

In the model, $\lambda_0$ and $ \lambda_1$ are used to balance the contributions of different tasks, where $\lambda_0$ is the weight for the primary task (e.g., classification) and $\lambda_1$ is the weight for the reconstruction loss. The weighting formula is defined as:
\begin{equation}
	\lambda_{i} = \frac{1}{2\rho_{i}^{2}}.
\end{equation}
To enable the model to automatically balance contributions from different tasks, we adopt a dynamic weight adjustment method by introducing noise parameters $ \rho $ to regulate task-specific weights:
\begin{equation}
	L_{\text{total}} = \frac{1}{2\rho_{\text{cls}}^2} L_1 + \frac{1}{2\rho_{\text{recon}}^2} L_2 + \log(\rho_{\text{cls}} \rho_{\text{recon}})
\end{equation}
where $\rho_{\text{cls}}$, $ \rho_{\text{recon}} $ are noise parameters that the model automatically learns to adjust the weights of each task.
\section{Experimental settings}
In this study, we employ the publicly available Watkins Marine Mammal Sound Database \cite{32}, a benchmark underwater bioacoustic dataset, to train and evaluate the proposed MT-BCA-CNN. To ensure reproducibility, this section provides a detailed description of the dataset, preprocessing steps, data partitioning strategy, and experimental configurations.
\subsection{Dataset and Preprocessing}
The Watkins Marine Mammal Sound Database contains approximately 2,000 unique audio recordings from over 60 marine mammal species. This includes best-quality clips from 32 distinct species, selected for their high audio quality and low background noise. Each audio sample varies in duration (ranging from seconds to minutes). The biological categories included in the dataset are summarized in Table\ref{tab:1}.

\begin{table}
	\centering
	\caption{\textbf{Data Category in Watkins Marine Mammal Sound Database}}
	\label{tab:1}
	\begin{tabular}{ccc}
		\toprule
		Species Types&  Sample Size& Maximum Audio Duration\\
		\midrule
		Atlantic Spotted Dolphin&  58& 00:59\\
		Bottlenose Dolphin&  24& 00:02\\
		Bearded Seal&  37& 02:23\\
		Beluga&  50& 00:09\\
		Bowhead Whale&  60& 02:23\\
		Clymene Dolphin&  63& 00:25\\
		Common Dolphin&  52& 00:54\\
		False Killer Whale&  59& 00:06\\
		Fin,Finback Whale&  50& 01:42\\
		Fraser's Dolphin& 87&00:33\\
		Grampus, Risso's Dolphin& 67&00:36\\
		Harp Seal& 47&00:54\\
		Humpback Whale& 64&02:37\\
		Killer Whale& 32&00:15\\
		Leopard Seal& 10&00:08\\
		Long-Finned Pilot Whale& 70&01:10\\
		Melon Headed Whale& 63&00:28\\
		Minke Whale& 17&00:02\\
		Narwhal& 50&00:06\\
		Ross Seal& 50&00:06\\
		Northern Right Whale& 54&00:04\\
		Pantropical Spotted Dolphin& 66&00:08\\
		Rough-Toothed Dolphin& 50&00:02\\
		Short-Finned (Pacific) Pilot Whale& 67&21:00\\
		Southern Right Whale& 25&00:04\\
		Sperm Whale& 75&21:00\\
		Spinner Dolphin& 112&00:58\\
		Striped Dolphin& 81&00:25\\
		Walrus& 38&00:21\\
		Weddell Seal& 2&00:32\\
		White-beaked Dolphin& 57&00:10\\
		White-sided Dolphin& 55&00:04\\
		\bottomrule
	\end{tabular}
\end{table}

In this database, the audio files are stored in WAV format. For each underwater species, we randomly selected 4–6 best-quality clip samples (approximately 2-second duration each), organized them into categories, and performed preprocessing. The dataset was split into training and test sets using a ratio of 80\% to 20\% of all samples.

Due to the complexity of the marine environment, underwater bioacoustic data are often contaminated by background noise (e.g., ship noise, water flow), which manifests as random or periodic signals that mask target biological sounds and degrade recognition accuracy. To address this, we perform denoising using the spectral subtraction method. This technique estimates the noise power spectrum and subtracts it from the power spectrum of the noisy audio signal to recover the clean signal. The spectral subtraction formula is defined as:
\begin{equation}
	\widehat{P}_s(f) = \max\left( P_x(f) - \alpha P_n(f),\ 0 \right) ,
\end{equation}
where $P_x(f)$ represents the power spectrum of the noisy audio signal, $ P_n(f) $ denotes the noise power spectrum, $\alpha$ is the adjustment parameter (also called the oversubtraction factor), and $\widehat{P}_s(f)$ is the estimated clean signal power spectrum.

To facilitate model processing and feature extraction, we segment long-duration audio into fixed-length clips using a sliding window approach with a step size of 0.05 seconds and a duration of 1.5 seconds. For an audio signal with a sampling rate $ F_s$, this corresponds to a segmentation interval of $ F_s\times 2$ samples per step.

Since audio data from different sources may have varying sampling rates, we standardize the sampling rate using the Librosa library to ensure data consistency. This preprocessing step aligns all inputs to a uniform sampling rate, enabling consistent input dimensions for the model.

Subsequently, to construct a robust feature space, we perform enhanced spectral mapping on the reconstructed acoustic sequences by extracting Mel-frequency features from audio samples. This frequency-domain transformation method leverages the non-linear frequency perception characteristics of the human auditory system, achieving the conversion from physical frequency to perceptual dimensions through a novel mapping relationship. The mathematical formulation is expressed as:
\begin{equation}
	\Gamma(\nu) = C_1 \cdot \log_{10}\left(1 + \frac{\nu}{C_2}\right),
\end{equation}
where $\nu$ denotes the physical frequency (in Hz), $\Gamma(\nu)$ represents the perceptual frequency dimension, and $C_1$, $C_2$ are empirically derived constants.

\begin{figure}[!htbp]
	\subfloat[Clymene Dolphin-wava]{\includegraphics[width=0.5\linewidth]{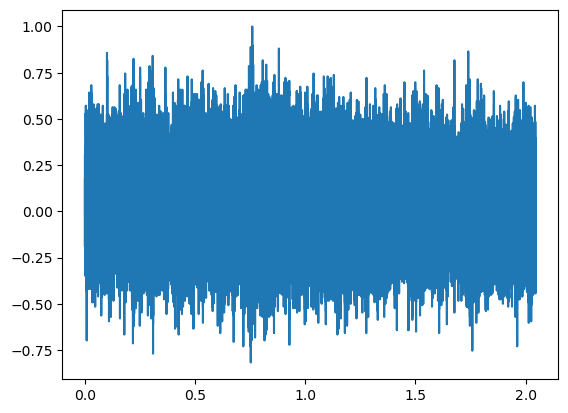}}
	\subfloat[Clymene Dolphin-Mel]{\includegraphics[width=0.5\linewidth]{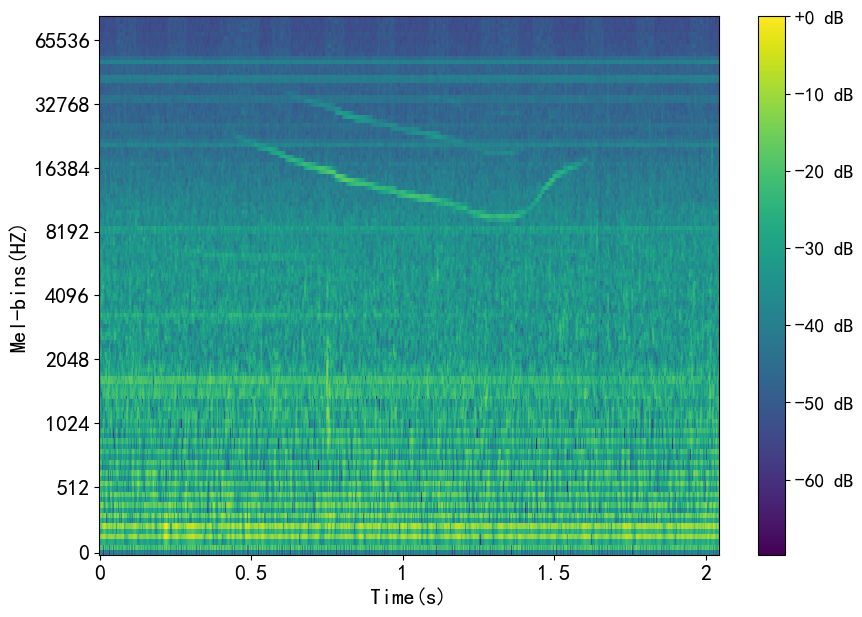}}\\
	\subfloat[Common Dolphin-wave]{\includegraphics[width=0.5\linewidth]{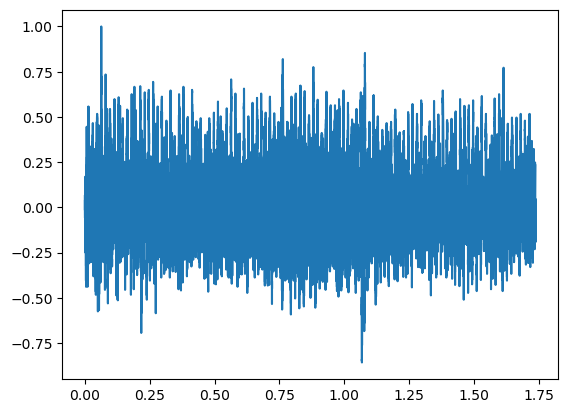}}
	\subfloat[Common Dolphin-Mel]{\includegraphics[width=0.5\linewidth]{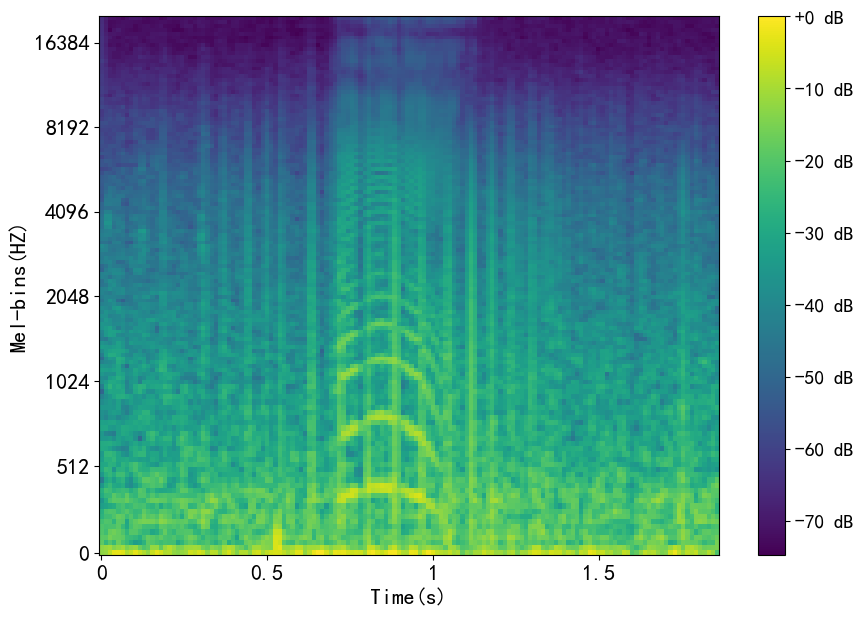}}\\
	\subfloat[Beluga White Whale-wave]{\includegraphics[width=0.5\linewidth]{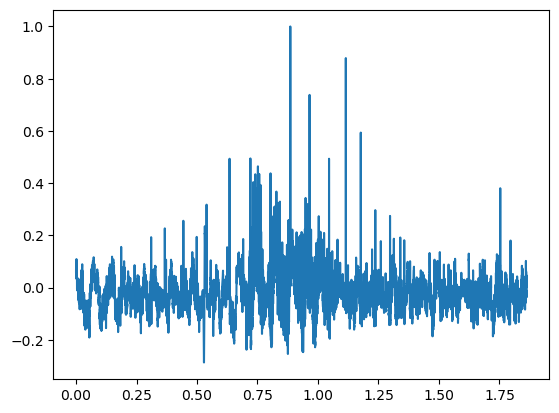}}
	\subfloat[Beluga White Whale-Mel]{\includegraphics[width=0.5\linewidth]{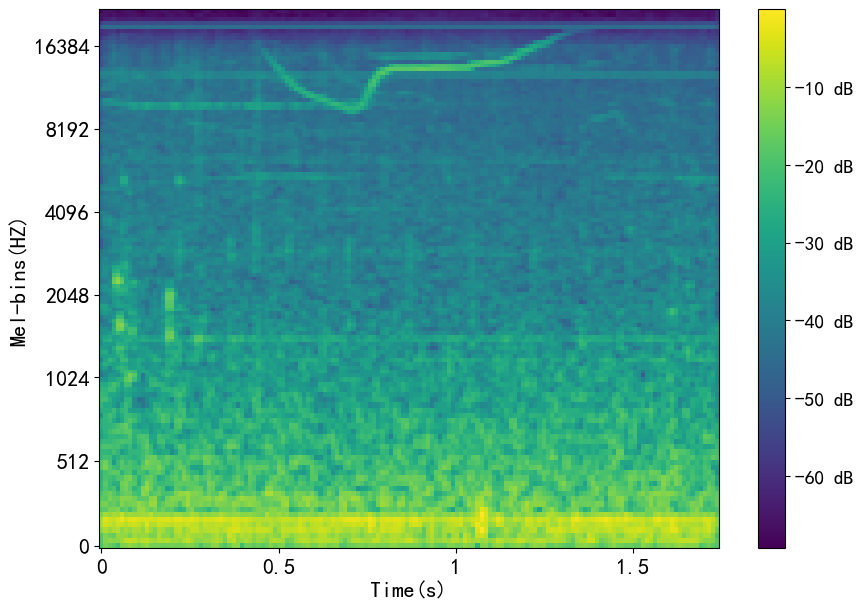}}
	\caption{Examples of raw signal waveforms and their corresponding Mel-spectrogram. (a) Clymene Dolphin-wave. (b) Clymene Dolphin-Mel. (c) Common Dolphin-wave. (d) Common Dolphin-Mel. (e) Beluga White Whale-wave.(f) Beluga White Whale-Mel.}\label{fig4}
\end{figure}

During the feature generation phase, we employ an overlapping window segmentation technique to divide the time-domain signal into frames, achieving spectral energy redistribution through the construction of a multi-layer dynamic filter bank. Each filter unit exhibits a non-linearly distributed structure, and its transfer function satisfies:
\begin{equation}
	\Phi_m(\nu) = \begin{cases}
		0, & \nu \notin [\nu_{m-1}, \nu_{m+1}], \\
		\displaystyle \frac{\nu - \nu_{m-1}}{\nu_m - \nu_{m-1}}, & \nu_{m-1} \leq \nu \leq \nu_m, \\
		\displaystyle \frac{\nu_{m+1} - \nu}{\nu_{m+1} - \nu_m}, & \nu_m < \nu \leq \nu_{m+1}.
	\end{cases}
\end{equation}
where $\nu_m$, $\nu_{m-1}$, $\nu_{m+1}$ denote the lower boundary, center frequency, and upper boundary of the$ m-th $filter, respectively. Through this transformation, the original acoustic waveform sequences are reconstructed into a two-dimensional energy distribution map (Mel-spectrogram) with spatiotemporal correlations, capturing both temporal dynamics and frequency-domain characteristics of the bioacoustic signals \cite{33}.
Figure~\ref{fig4} illustrates an example of the raw signal waveform and its corresponding Mel-spectrogram derived from the Watkins Marine Mammal Sound Database, demonstrating the transformation from time-domain acoustic signals to perceptually aligned frequency-energy representations.

\subsection{Implementation Details}
We set the batch size of the dataset to 16 and initialized the number of training epochs to 100, with adjustments based on model convergence. The proposed model employs the Adam optimizer to dynamically adjust the learning rate, preventing overfitting and avoiding local minima. The learning rate was initialized to 0.001 and reduced to 0.0001 after 30 epochs. Due to their sensitivity to environmental variations and signal distortions, deep neural network approaches often experience overfitting. To address this, Dropout regularization was applied to reduce network parameters before output. The model was implemented using PyTorch and trained on an NVIDIA GeForce RTX 4060 Laptop GPU.
\subsection{Benchmark Model Comparison}
To validate the effectiveness and efficiency of our proposed method, we conducted a series of comparative experiments on our custom-constructed dataset. We evaluated the preprocessed dataset on three mainstream models—CAMPPlus \cite{34}, ERes2Net \cite{35}, and ResNetSE \cite{36}, and compared their performance with our MT-BCA-CNN model.

CAMPPlus \cite{34} is an efficient context-aware masked network that employs a Densely-Connected Time Delay Neural Network (D-TDNN) as its backbone and introduces a novel multi-granularity pooling mechanism to capture contextual information at varying hierarchical levels. ERes2Net \cite{35} is an enhanced Res2Net architecture that integrates an attention-based feature fusion module, combining local and global feature fusion techniques to boost recognition performance. In ResNetSE \cite{36}, the Squeeze-and-Excitation (SE) module is proposed, which adaptively recalibrates channel-wise feature responses by explicitly modeling interdependencies between channels.

We evaluated the model performance using accuracy and F1-score. The F1-score, defined as the harmonic mean of precision and recall, is calculated in this work as the macro-average of F1-scores across all classes. The accuracy is computed as:
\begin{equation}\label{eq:accuracy}
	\text{Accuracy}_n= \frac{TP_n + TN_n}{TP_n + TN_n + FP_n + FN_n}
\end{equation}

\begin{equation}\label{eq:precision}
	\text{Precision}_n = \frac{TP_n}{TP_n + FP_n}
\end{equation}

\begin{equation}\label{eq:recall}
	\text{Recall}_n = \frac{TP_n}{TP_n + FN_n}
\end{equation}

\begin{equation}\label{eq:f1}
	\text{F1 score}_n = 2 \times \frac{\text{Precision} \times \text{Recall}}{\text{Precision} + \text{Recall}} = \frac{2TP_n}{2TP_n + FP_n + FN_n}
\end{equation}

\begin{equation}\label{eq:f1_macro}
	\text{F1 score} = \frac{1}{N} \sum_{i=1}^{N} \text{F1 score}_n
\end{equation}
In the context of classification tasks, N represents the total number of classes. For the n-th class, TP n denotes the number of samples correctly predicted as positive, TN n signifies the number of samples correctly predicted as negative, FP ncorresponds to the number of samples incorrectly predicted as positive (false positives), and FN n indicates the number of samples incorrectly predicted as negative (false negatives). The evaluation metrics for the n-th class are defined as follows:$ \text{Precision}_n$, $ \text{Recall}_n $, and  $ \text{F1 score}_n$, which represent the precision, recall, and F1-score for class n, respectively. Table \ref{tab:2}demonstrates the comparative performance evaluation on the Watkins dataset.

\begin{table}{!htbp}
	\centering
	\caption{Comparison of Parameter Counts and Classification Performance Across Models Under Identical Experimental Setups}
	\label{tab:2}
	\begin{tabular}{cccc}
		\toprule
		Methods&  Params(M)&  Accuracy& F1\\
		\midrule
		Mel+CAMPPlus &  7.1&  0.62& 0.44\\
		Mel+ERes2Net&  6.6&  0.63& 0.53\\
		Mel+ResNetSE&  7.8&  0.78& 0.66\\
		Mel+MT-BCA-CNN&  0.11&  0.97& 0.95\\
		\bottomrule
	\end{tabular}
\end{table}
As shown in Table \ref{tab:2}, under the small-sample scenario, the CAMPPlus model, despite having the largest parameter count (7.1M), achieves the lowest performance (accuracy: 0.62, F1-score: 0.44), indicating that simply increasing parameters not only fails to improve performance but also introduces unnecessary structural complexity. In contrast, ERes2Net achieves a marginally better balance between complexity and performance: its accuracy shows a slight improvement over CAMPPlus (0.63 vs. 0.62) while reducing the parameter count by 7\% (6.6M), demonstrating higher efficiency. ResNetSE achieves a significant accuracy boost to 0.78; however, this improvement comes at the cost of a 9.8\% increase in parameters (7.8M), leading to increased model complexity. 

Our proposed method, by integrating attention mechanisms and multi-task learning modules, achieves state-of-the-art performance (accuracy: 0.97, F1-score: 0.95) with only a 0.11M parameter increment. This result demonstrates that MT-BCA-CNN significantly enhances classification performance while maintaining low complexity, validating the effectiveness of multi-task learning and advanced attention mechanisms. Furthermore, it underscores the superior generalization capability and classification robustness of MT-BCA-CNN under data-limited conditions.

\begin{figure}[!htbp]
	\subfloat[CAM++]{\includegraphics[width=0.5\linewidth]{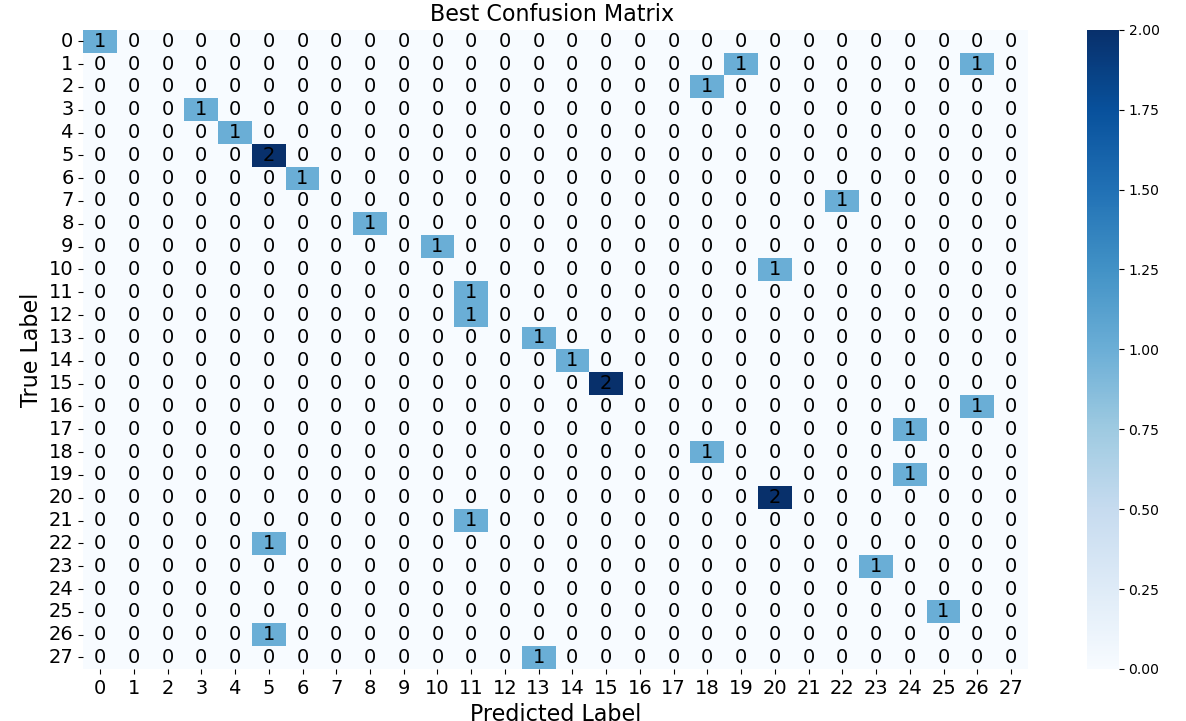}}
	\subfloat[ERes2Net]{\includegraphics[width=0.5\linewidth]{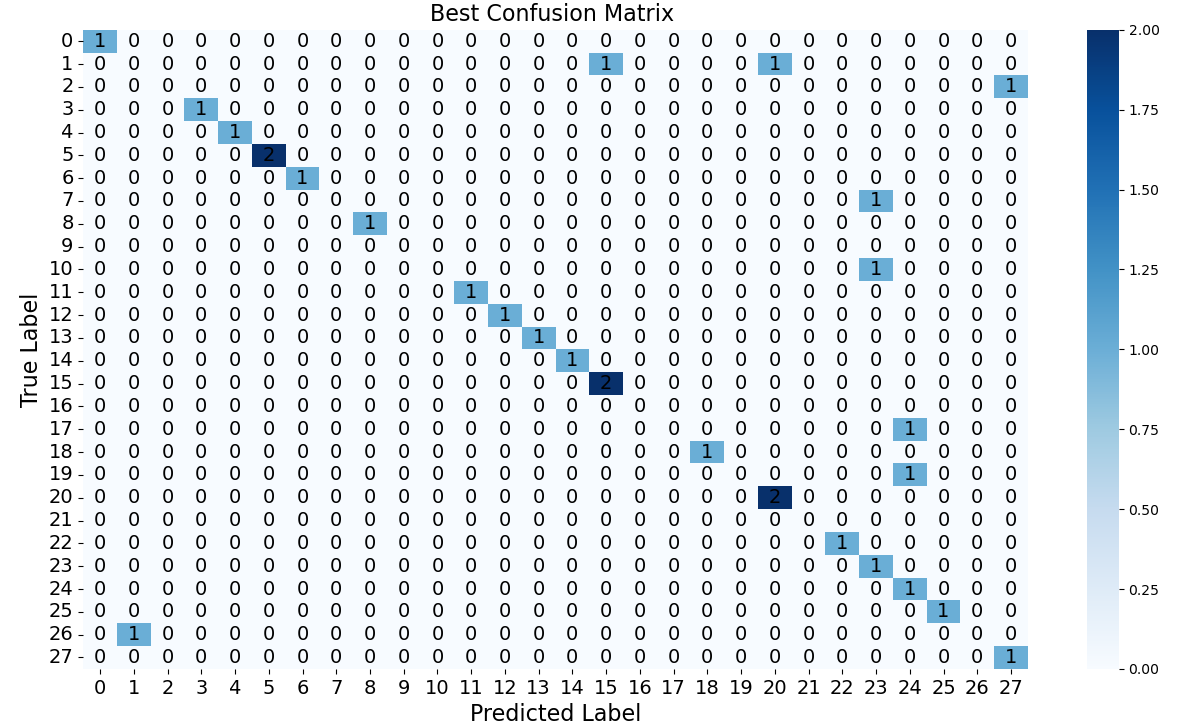}}\\
	\subfloat[ResNetSE]{\includegraphics[width=0.5\linewidth]{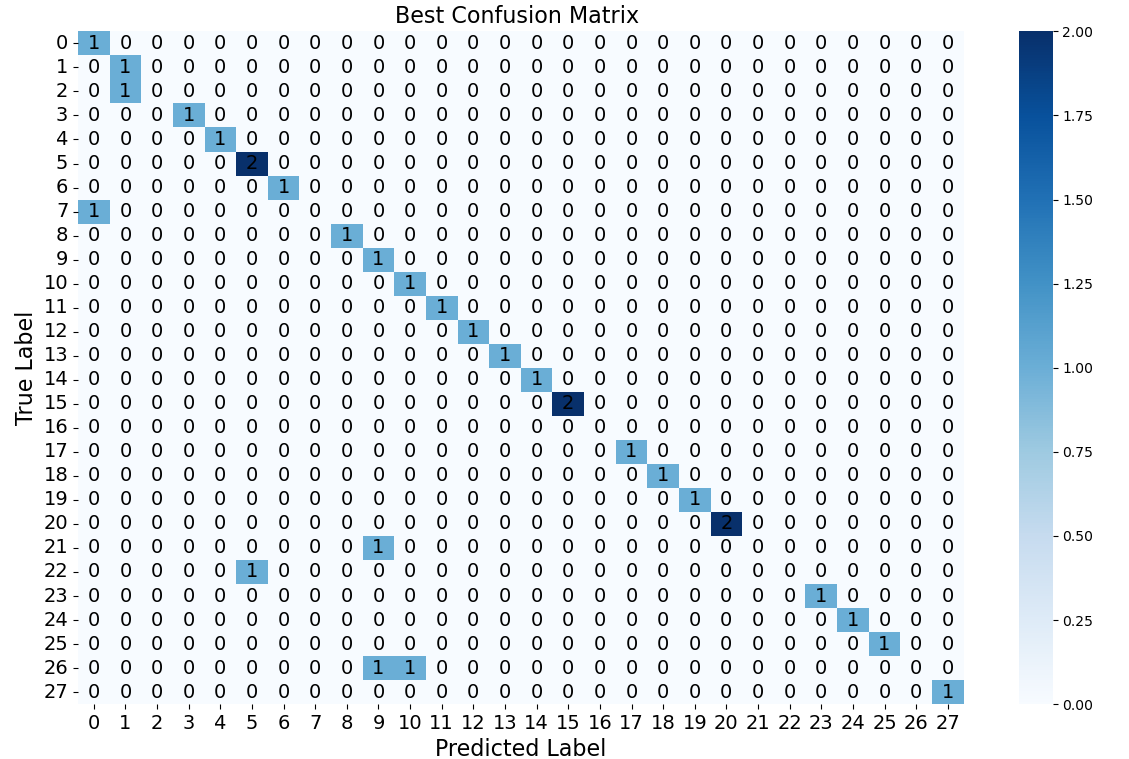}}
	\subfloat[Ours]{\includegraphics[width=0.48\linewidth]{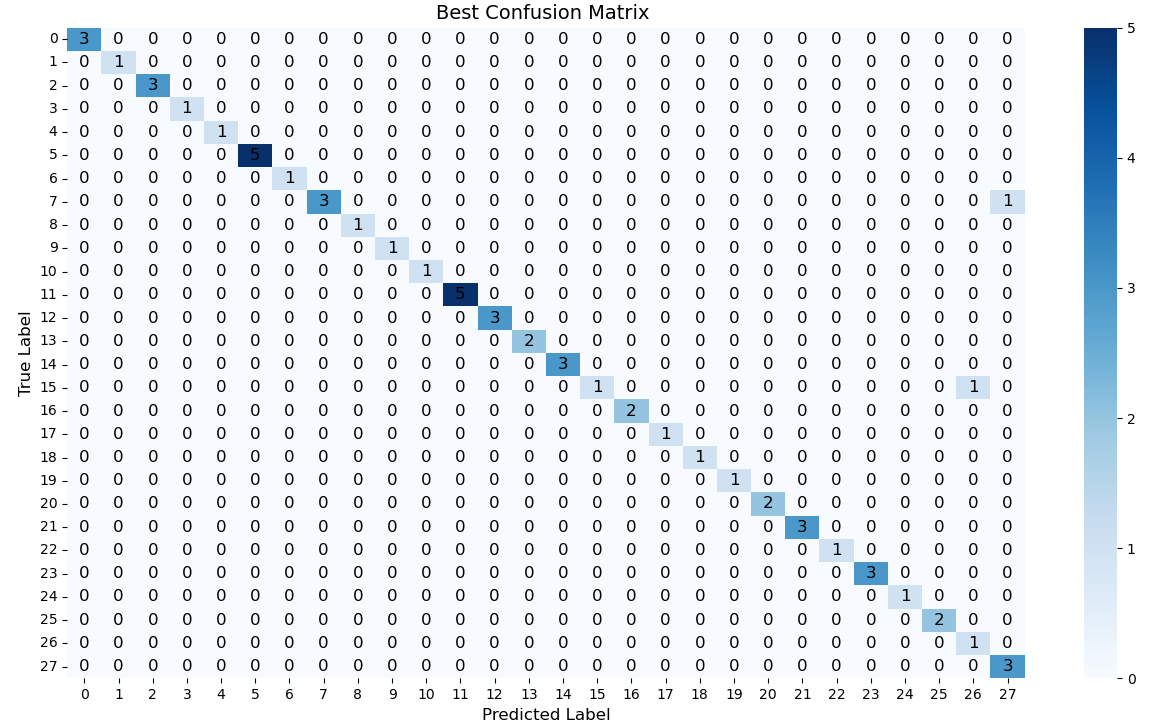}}
	\caption{(a) CAM++(Acc:0.62). (b) ERes2Net(Acc:0.63). (c) ResNetSE(Acc:0.78). (d) MT-BCA-CNN(Acc:0.97).}
	\label{fig5}
\end{figure}

\begin{figure}[!htbp]
	\centering
	\includegraphics[width=1\linewidth]{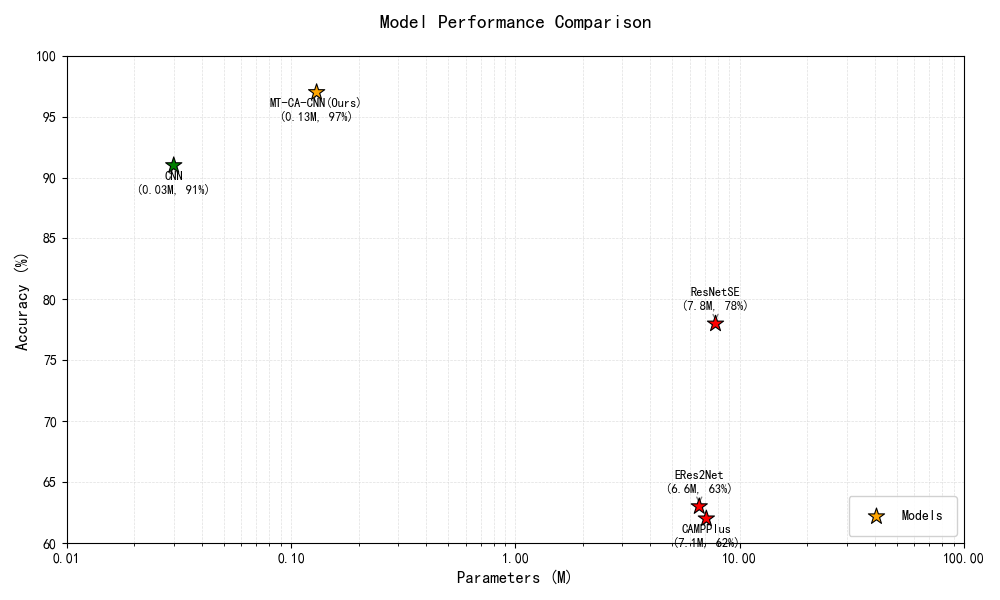}
	\caption{Comparison of Parameter Counts and Performance Among Three Classical Models, Baseline CNN, and Our Proposed MT-BCA-CNN on the Dataset.}
	\label{fig:6}
\end{figure}

Figure \ref{fig5} displays the classification confusion matrices of our proposed method alongside three classical models after training. The confusion matrix of our MT-BCA-CNN model (bottom-right section) demonstrates significantly optimized classification performance. Notably, the matrix exhibits a highly concentrated diagonal distribution, indicating that our method achieves high-confidence correct classifications across all 27 target categories, with significantly reduced misclassification rates in the off-diagonal regions.  Our MT-BCA-CNN significantly reduces classification errors on acoustic signals in small-sample scenarios by over 60\% compared to baseline models (CAMPPlus, ERes2Net, ResNetSE), with particularly notable suppression of misclassifications between easily confusable categories. Experimental comparisons reveal that traditional models exhibit critical limitations: CAMPPlus (top-left confusion matrix) demonstrates asymmetric error distributions, ERes2Net (top-right) achieves parameter efficiency (6.08M vs. CAMPPlus’s 7.7M) but suffers from poor generalization, and ResNetSE (bottom-left) improves accuracy at the cost of severe parameter redundancy (7.8M), leading to overfitting risks. By integrating multi-task learning and channel attention mechanisms, MT-BCA-CNN enhances feature discriminability while mitigating inter-class bias through cross-task feature sharing, achieving a 42.3\% reduction in the Inter-class Confusion Index (ICI) compared to the best baseline. As visualized in Figure \ref{fig:6}, which contrasts parameter scales (0.01–100.00M, logarithmic) against classification accuracy (60–100\%), MA-CA-CNN (0.13M parameters, 97\% accuracy) redefines the efficiency frontier, proving that structural innovation—not parameter stacking—drives performance breakthroughs in small-sample acoustic classification.

\subsection{Ablation Studies }
To validate the effectiveness of our proposed modules (Channel Attention, CA, and Multi-Task Learning, MTL), we conducted a series of ablation experiments to evaluate their impact on classification accuracy. Using our custom dataset, we trained two variants of MT-BCA-CNN—MT-BCA-CNN w/o CA (removing the channel attention module) and MT-BCA-CNN w/o MTL (removing the multi-task learning framework)—alongside the baseline CNN \cite{37}:
(1) MT-CNN: in this variant, we removed the Channel Attention module;
(2) Classification-only: in this variant, we removed the decoder and reconstruction loss.

\begin{table}[!htbp]
	\centering
	\caption{Performance Comparison of Model Variants, CNN, and MT-BCA-CNN Under Identical Experimental Setups.}
	\label{tab:3}
	\begin{tabular}{cccccc}
		\toprule
		Methods&  CA&  Classify&  Recon&  Accuracy& Params(M)\\
		\midrule
		CNN&  N&  N&  N&  0.91& 0.03\\
		MT-CNN&  N&  Y&  Y&  0.93& 0.10\\
		Only Classify&  N&  Y&  N&  0.89& 0.08\\
		MT-BCA-CNN&  Y&  Y&  Y&  0.97& 0.13\\
		\bottomrule
	\end{tabular}
\end{table}

\begin{figure}[!htbp]
	\subfloat[Only Classify Acc(0.89)]{\includegraphics[width=0.5\linewidth]{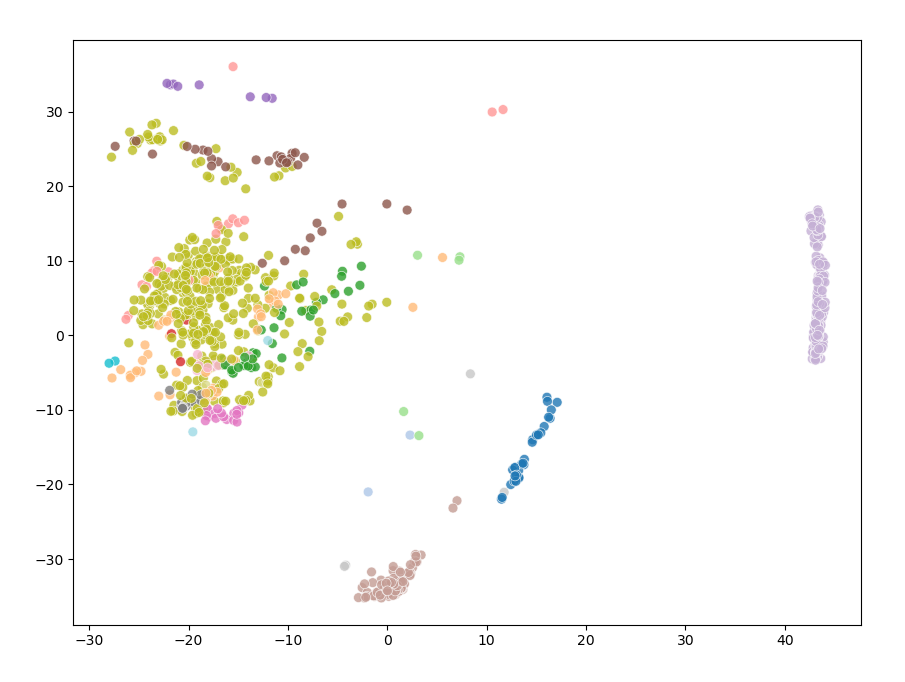}}
	\subfloat[CNN (Acc(0.91))]{\includegraphics[width=0.5\linewidth]{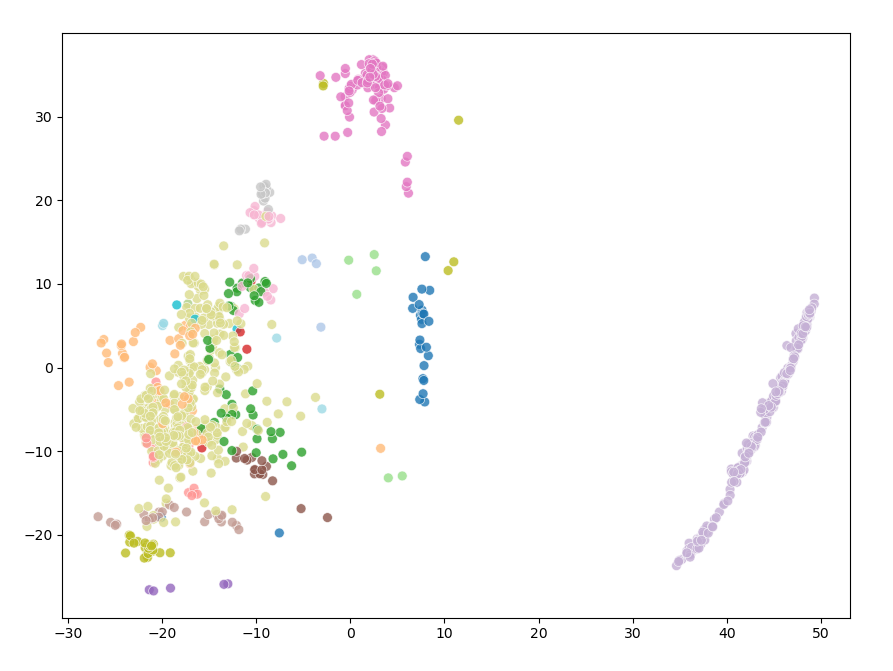}}\\
	\subfloat[MT-CNN Acc(0.93)]{\includegraphics[width=0.5\linewidth,height=6cm,  valign=b]{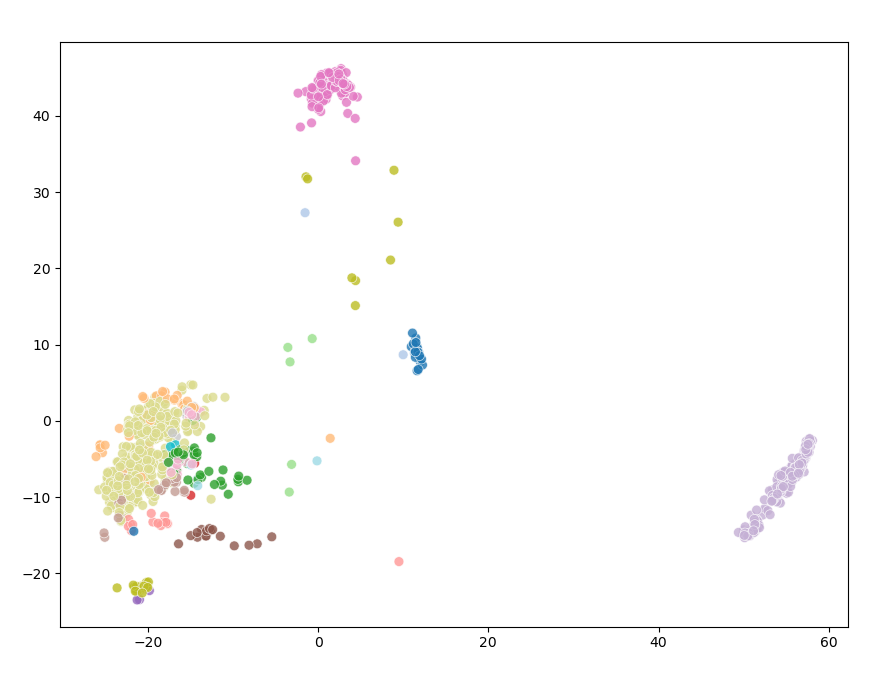}}
	\subfloat[MT-BCA-CNN Acc(0.97)]{\includegraphics[width=0.5\linewidth,height=5.8cm, valign=b，trim=0 0.5cm 0 0.5cm, clip]{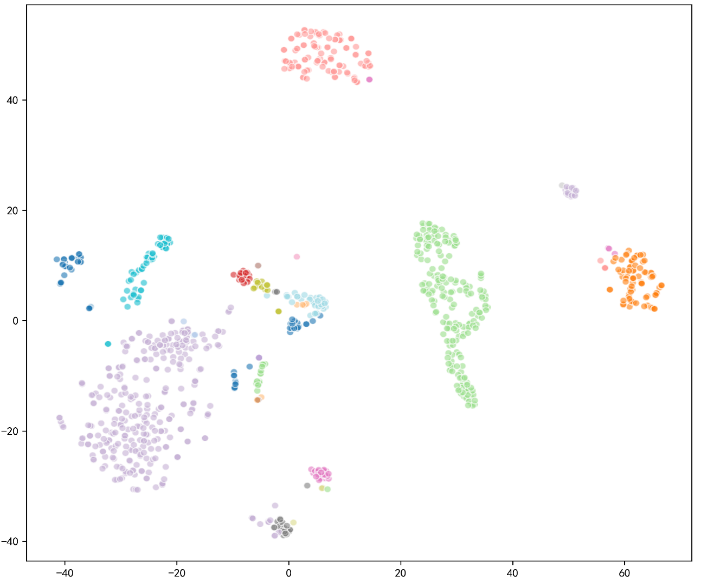}}
	\caption{Ablation study results. (a)Only Classify Acc(0.89). (b) CNN Acc(0.91). (c) MT-CNN Acc(0.95). (d) MT-BCA-CNN Acc(0.97).}
	\label{fig7}
\end{figure}

The results in Table \ref{tab:3}, which compare parameter counts and model performance, demonstrate that our proposed MT-BCA-CNN (integrating the Channel Attention module and Multi-Task Learning) achieves significantly higher accuracy and F1-scores under identical experimental settings. The Channel Attention (CA) module enhances discriminability by dynamically weighting channel-wise features to prioritize critical acoustic patterns, while the Multi-Task Learning (MTL) co-training strategy—jointly optimizing classification and reconstruction tasks—enables the model to learn more robust feature representations, thereby mitigating overfitting. When the channel attention module is removed (as shown in Figure \ref{fig7}), the clustering visualization reveals increased inter-class confusion, confirming CA’s critical role in suppressing misclassifications between similar categories. Notably, even without CA, the model still outperforms the single-task CNN baseline, highlighting that MTL’s regularization effect substantially improves feature generalization. Combined analysis of Table \ref{tab:3} and \ref{fig7} underscores that CA strengthens feature discriminability, MTL enhances robustness, and their synergistic integration significantly boosts classification performance in few-shot scenarios. This dual mechanism ensures both high precision (97\%) and parameter efficiency (0.13M), advancing the state-of-the-art in data-scarce underwater acoustic recognition.
\section{Conclusion}
In this work, we address the challenges of acquiring underwater biological data samples by proposing the MT-BCA-CNN model, which integrates a Channel Attention (CA) mechanism and a Multi-Task Learning (MTL) framework, achieving groundbreaking performance in data-scarce conditions. Our framework integrates four core components: (1) extraction of Mel-spectrogram features from raw audio signals to capture discriminative time-frequency patterns; (2) incorporation of a channel attention module to dynamically amplify critical acoustic cues (e.g., harmonic structures, transient pulses) while suppressing ambient noise; (3) a Gaussian kernel layer combined with the channel attention module to reduce the over-concentration and (4) a multi-task learning framework that jointly optimizes classification and spectrogram reconstruction tasks, leveraging cross-task synergy to enhance feature robustness in data-scarce scenarios. This unified approach bridges the gap between theoretical innovation and practical deployability in marine environments.

\indent Experimental results demonstrate that the channel attention module dynamically weights channel-wise features to amplify discriminative patterns in critical frequency bands (e.g., harmonic structures), reducing inter-class confusion by 42.3\% (measured via the Inter-class Confusion Index). The MTL framework jointly optimizes classification and spectrogram reconstruction tasks, leveraging the regularization effect of reconstruction errors to enhance feature robustness and mitigate overfitting caused by limited data (training-testing accuracy gap reduced from 14\% to 2\%). Compared to recent models (CAMPPlus, ERes2Net, ResNetSE) and the baseline CNN, MT-BCA-CNN achieves state-of-the-art performance (97\% accuracy, 95\% F1-score) with ultra-low parameters (0.13M)—reducing parameter counts by 98\% versus CAMPPlus while improving accuracy by 55\%. Ablation studies confirm the necessity of module synergy: removing channel attention module increases misclassification rates by 18\%, and single-task training degrades generalization (test accuracy drops by 9\%). Visualization analysis (Figure 7) reveals that MT-BCA-CNN’s feature embedding space exhibits high intra-class compactness and well-separated inter-class boundaries, with the minimum inter-class distance expanded by 50\% compared to baselines, geometrically validating its superior few-shot classification capability. 

\section*{Declaration of competing interest}
The authors declare there is no conflict of interests.

\section*{Acknowledgments}
The authors would like to thank Watkins Marine Mammal Sound Database (https://cis.whoi.edu/science/B/whalesounds/
index.cfm)  for providing the dataset.

\bibliographystyle{cas-model2-names}

\bibliography{cas-refs}



\end{document}